\documentclass[10pt,a4paper]{article}

\usepackage{blindtext} 

\usepackage[sc]{mathpazo} 
\usepackage[T1]{fontenc} 
\usepackage{microtype} 
\usepackage{tabularx}

\usepackage[english]{babel} 

\usepackage[top=3cm,bottom=3cm,left=2cm,right=2cm]{geometry}

\usepackage{booktabs} 

\usepackage{enumitem} 
\setlist[itemize]{noitemsep} 

\usepackage{abstract} 

\usepackage{titlesec} 
\titleformat{\section}[block]{\large\scshape\centering}{\thesection.}{1em}{} 
\titleformat{\subsection}[block]{\large}{\thesubsection.}{1em}{} 

\usepackage{fancyhdr} 
\usepackage{lmodern,textcomp}
\usepackage{titling} 

\usepackage{hyperref} 
\usepackage{graphicx}

\pretitle{\begin{center}\Huge\bfseries} 
\posttitle{\end{center}} 
\title{How to automate a kinematic mount using a 3D printed Arduino-based system} 

\author{%
\textsc{Luis Jos\'{e} Salazar-Serrano$^{1}$, Gerard Jim\'enez$^{2}$ and Juan P. Torres$^{2,3}$} \\[1ex] 
\normalsize $^{1}$Aistech Space, Esteve Terrades 1, Despacho 110, 08860, Castelldefels, Barcelona, Spain\\
\normalsize $^{2}$ICFO - The Institute of Photonic Sciences, the Barcelona Institute of Science and Technology\\ \normalsize 08860, Castelldefels, Barcelona, Spain \\ \normalsize $^{3}$Dep. Signal Theory and Communications, Universitat Politecnica de Catalunya\\ \normalsize Campus Nord D3, 08034 Barcelona, Spain \\
\normalsize Corresponding author: ljsalazarserrano@gmail.com 
}
\date{\today} 

\renewcommand{\maketitlehookd}{%
\begin{abstract}
\normalfont{We demonstrate a simple, flexible and cost-effective
system to automatize most of the kinematic mounts available
nowadays on the market. It combines 3D printed components, an
Arduino board, stepper motors and simple electronics.  The system
developed can control independently and simultaneously up to ten
stepper motors using commands sent through the serial port, and it
is suitable for applications where optical realignment using flat
mirrors is required on a periodic basis.}
\end{abstract}
}

\begin{document}
\renewcommand{\abstractname}{\vspace{-\baselineskip}}

\maketitle




\section{Introduction}
Over the last years the combination of 3D printing and Arduino has
proved to be an efficient and cost-effective way of making
experimental equipment, becoming in many occasions
an attractive alternative to its commercial counterparts. This
combination of Arduino, 3D printed parts and simple hardware have
already been used in a system to determine the concentration of
dissolved species \cite{22}, to provide a carefully controlled
dose of a given reagent \cite{24}, to characterize easily the
spatial profile of a laser beam \cite{26}, and to centrifuge a
substance for DNA extraction \cite{23}, among others.

In all of these cases, expensive and sometimes over-engineered
components are substituted by equipment that, despite costing a
fraction of its commercial counterpart, can be used to perform
state-of-the-art research. Moreover, thanks to the fact that the
fabrication time is minimal, the characteristics and quality of
the components evolve very fast in the hands of researchers. This
evolution process has unexpectedly turned out to be a creative way
to engage young minds to research in natural sciences.

In this paper we report the development of a system that
can motorize many of the kinematic mounts
available nowadays on the market. It combines 3D printed
components, simple hardware and electronics, and an Arduino board.
The system is intended to add automation capabilities to the
Open-Source Optics Library \cite{8} already available on the
Internet. The library is composed of a broad selection of optical
components ubiquitous in any optics experiment such as lens
holders \cite{9}, screen/filter holders \cite{10, 11}, lab jacks
\cite{12}, fiber optic holders \cite{13}, kinematic mirrors and
translation stages \cite{14, myplos}, and parametric open-source
chopper wheels \cite{15}, to name a few.

\begin{figure}[!ht]
\centering
\includegraphics[width=6cm]{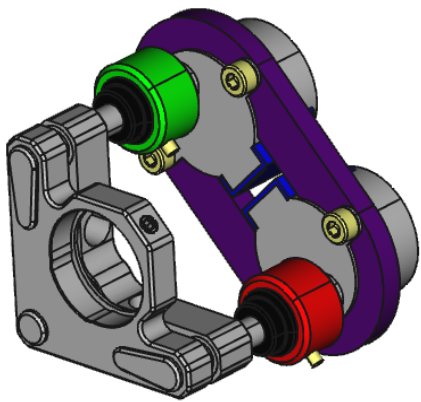}
\caption{Design in Freecad showing the implementation of the motorized system for a kinematic mount from
Thorlabs (KM100). 3D printed components are shown in green, red and purple.} \label{fig:figure1}
\end{figure}

The system we demonstrate here is simple, flexible and
cost-effective. The simplicity of the system relies on the
components used to implement it. It is composed of a 3D printed
plate that supports two stepper motors 28BYJ-48. Its shafts are
aligned with the knobs used to tip and tilt the plane of the
kinematic mount that supports the optics (see Figure
\ref{fig:figure1}). The cylindrical adapter used to couple the
shaft to the knob is also 3D printed. An Arduino MEGA controls the
motors. In order to simplify the connections with the motor
drivers, a shield (Sensor Shield V1.0) provides the capability of
controlling up to ten motors simultaneously by sending simple
instructions defined by a command table through the serial port.

The hardware and software of the system are flexible and can be
easily modified according to the user needs. From the hardware
side, since the 3D printed components that couple the system to
the kinematic mount were parametrically designed using FreeCAD
0.16 \cite{freecad}, the system is highly tunable by changing
only a few parameters on a spreadsheet. Similarly
from the software side, the Arduino code is well documented and
some usage examples are provided in the {\em *.ino} file. The
FreeCAD and Arduino source code can be downloaded from the
repositories Thingiverse \cite{thingiverse} and Github
\cite{github}, respectively.

The low cost of the system is the byproduct of its simplicity. As
a reference, the average cost of fabricating a system composed of
four stepper motors for controlling two kinematic mounts is around
30 EUR . The system presented in this letter compares thus
favorably with respect to commercial alternatives, since it can be
roughly between 5 to 10 times less expensive. However, a word of
caution should be added here. Since the system keeps track of the
number of steps given by the stepper motor of each channel to
estimate its current position, some inaccuracies may appear due to
the hysteresis present on the stepper motors. Therefore, the {\em
true} usefulness of the system presented in this work will depend
on its specific aim. Based on our experience in the laboratory in
a research institute, we find that the system is suitable for a
wealth of applications ranging from teaching to doing high-quality
research experiments.

To illustrate and validate the system developed, we consider two
scenarios that may require the use of two mirrors supported by two
kinematic mounts. The first scenario is the alignment of a laser
beam through two reference apertures, and the second scenario is
the coupling of light into a multimode or single-mode optical
fiber. Both scenarios might be of great interest to experimenters
that require to re-align an optical beam or to increase the
coupling efficiency of an optical fiber on a daily basis.

\section{Results}
\begin{figure}[!ht]
\centering
\includegraphics[width=12cm]{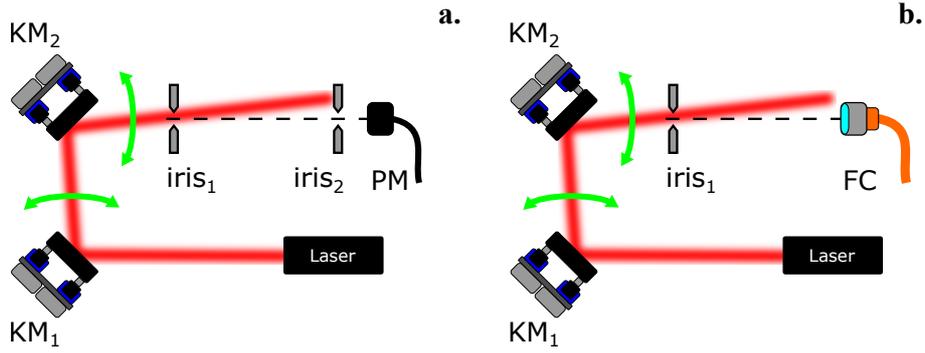}
\caption{Experimental schemes. $\mathrm{KM_1}$ and
$\mathrm{KM_2}$: Kinematic mounts; PM: Power meter; FC: Fiber
coupler.} \label{fig:figure2}
\end{figure}

To validate and test the performance of the motorized system, we
have implemented the two experimental schemes shown in Figure
\ref{fig:figure2}. We use a HeNe laser, and align the laser beam
with the help of two mirrors supported by two kinematic mounts,
$\mathrm{KM_1}$ and $\mathrm{KM_2}$.

In each experiment, the tip and tilt of each of the two mirrors is
automatically controlled using a MATLAB routine. The power after
the second iris (Figure \ref{fig:figure2}(a)) or at the output of
the fiber (Figure \ref{fig:figure2}(b)) is monitored using a
Thorlabs PM100A power meter. After scanning sequentially the tip
and subsequently the tilt of both mirrors, the final position of
each kinematic mount is set to the position that corresponds to
the maximum power registered.

\subsection{Alignment of an optical beam with the help of two irises}
The experimental scheme is shown in Figure \ref{fig:figure2}(a).
The initial position of the mirrors (without using the stepper
motors) is set so that the optical beam hits the first iris close
to its center. This coarse alignment is mainly driven by the first
mirror $\mathrm{KM_1}$. Once the starting point is defined, the
motors are attached to the kinematic mounts knobs and a simple
MATLAB routine is executed where each motor move independently a
defined number of steps either in a clock-wise or anti-clockwise
direction. The optical power is measured by the detector located
after the second iris.

\begin{figure}[!ht]
\centering
\includegraphics[width=12cm]{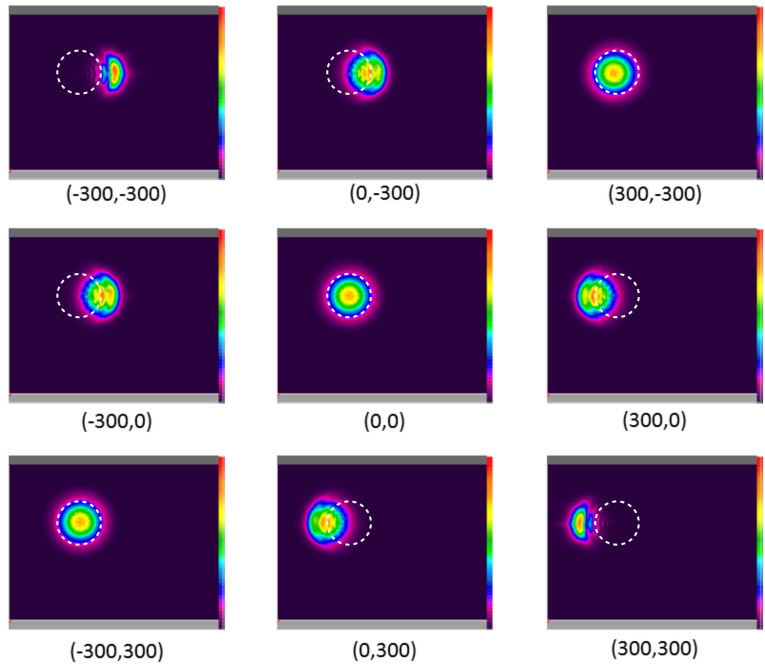}
\caption{X-direction scan. Below each capture, in brackets,
(X1-Motor Position, X2-Motor Position).} \label{fig:figure3}
\end{figure}

\begin{figure}[!ht]
\centering
\includegraphics[width=12cm]{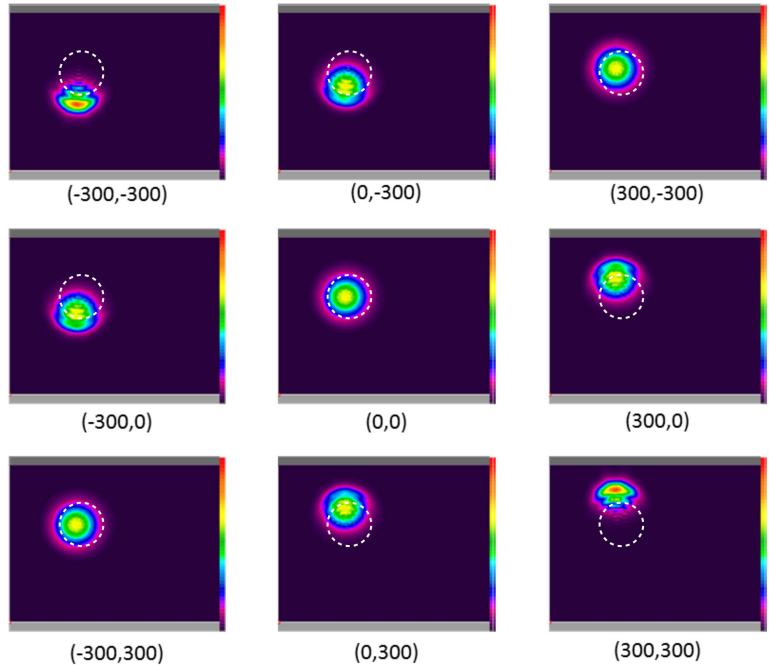}
\caption{Y-direction scan. Below each capture, in brackets,
(Y1-Motor Position, Y2-Motor Position).} \label{fig:figure4}
\end{figure}

To illustrate a step by step run of the MATLAB routine, a CCD
camera is placed just after the second iris and a beam image is
acquired for several positions of both motors. Figures
\ref{fig:figure3} and \ref{fig:figure4} show the beam position
registered for a given combination of steps $(N_1,N_2)$ in the X
and Y directions, respectively. As a reference, the position of
the second iris (set at a given aperture) is indicated by a dashed
circle in all images.

In the example, each motor $KM_1$ and $KM_2$ is set to move
between three different positions separated by 300 steps, so that
each motor can move between the absolute positions -300, 0 and
300. As a result, a full scan in the X and Y
directions is defined by 9 different positions
indicated by the array $(N_1, N_2)$ below each figure.

Notice that the maximum power is registered for the positions
$(0,0)$, while the positions with opposite angles $(-300,+300)$
and $(+300,-300)$ correspond to a situation where the effect of
one mirror is compensated by the other and the centroid of each
beam is slightly shifted to the right or left, respectively.

\subsection{Optimizing the coupling efficiency of a light beam into an optical fiber}
We replace the second iris of the previous experimental setup by a
multi-mode optical fiber (see Figure \ref{fig:figure2}(b)). The
aim is to test the performance of the system, and the simple
routine written in MATLAB code, for optimizing the coupling of
light into an optical fiber. The motors attached to the 3D printed
kinematic mounts perform a scan in the X direction, and later on
in the Y direction, controlled by a MATLAB program.

\begin{figure}[!ht]
\centering
\includegraphics[width=14cm]{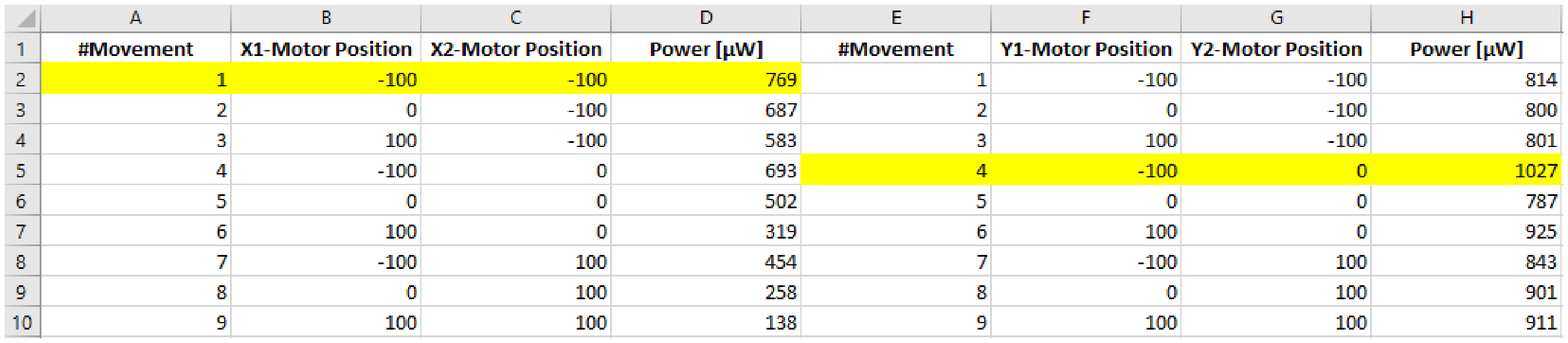}
\caption{Optimization of light coupling into an optical fiber. Example of the data
collected and saved in the \textit{.dat} file generated by the
MATLAB program after performing a scan in the X and Y directions.}
\label{fig:table}
\end{figure}

Figure \ref{fig:table} shows a table with the data obtained after
performing a full scan in both directions. First the X scan
provides a power maximum (highlighted), which is used as the
starting point for a second scan in the Y direction (highlighted).

After the scan in the X and Y directions the motors will move to
positions $(-100,-100)$ and $(-100,0)$ respectively. For this
final situation, the recorded power is 1027 $\mu W$. Taking into
account that the input power of the laser is 1250 $\mu W$,
we obtain that the laser-fiber coupling
efficiency is $82.16 \%$.

\section{Discussion}
The combination of 3D printing and simple hardware and electronics
components is becoming more and more popular as a tool for
developing scientific equipment. It offers the possibility to
lower significantly the cost of an experiment, and sometimes even
more importantly, it helps to reduce dramatically the delivery
time of certain components.

\begin{table}[h!]
\centering
\footnotesize
\begin{tabularx}{\textwidth}{X X X X}
\toprule
\textbf{Company}    & \textbf{Software} & \textbf{Hardware} & \textbf{Price [ EUR ]} \\
\midrule
Newport & Picomotor\textsuperscript{TM} & 8806 Picomotor Motorized Blank Plate Mount &  1885\\
Edmund Optics & Via RS-232 Serial Interface & Motorized Kinematic Mount &  1595\\
Standa & 8SMC5-USB Stepper and DC Motor Controller & 8MBM24 - Motorized Mirror Mounts &  940 + 699\\
Physik Instrumente (PI) & PI Mikromove & N-480 PiezoMike Linear Actuator & 2109.03\\
Thorlabs & K-Cube DC Servo Motor Controller & KS1-Z8 DC Motorized Mirror Mount &  580.41 + 1218.97\\
\bottomrule
\end{tabularx}
\caption{Selected commercially
available motorized kinematic mounts.} \label{tab:commercial}
\end{table}

One important conclusion that is extracted from Table
\ref{tab:commercial} is that worldwide there are only a few
companies that sell optical equipment that cover satisfactorily
the basic need to automate the alignment of optical systems. In
any case, prices are reasonably high, that leads to the need to
reduce them using new resources. Precisely our work addresses this
last concern, and makes use of 3D printing and simple electronics to lower cost.

Using 3D printed kinematic mounts dramatically reduces the cost,
and greatly increases flexibility and adaptability with respect to
existing systems. The experimental results obtained support the
operation and viability of both the low-cost hardware and the
simple open-source software provided.

The convergence of the auto-alignment method is undoubtedly key
and essential to ensure its proper functioning. In our case, this
convergence is guaranteed by reducing the number of steps of the
motors in the feedback loop. In other words, the convergence must
be done by the user, who must manually reduce the number of steps.
For the same reason, an obvious improvement of our MATLAB routine
would be the "automatization" of that convergence, that is, a code
designed to converge to the optimal solution that reduces the
number of steps of the motors automatically.

Another possible improvement that can be implemented in our code
would be the realization of a single feedback loop for the X-Y
plane scan. Recall that in our case there are two loops, one for
scanning along the X direction and
another one for the Y direction. This seems
reasonable, since it simulates the alignment that is carried out
manually by scientists. But for greater compactness and lower
computational cost, a single loop that synchronizes the movement
of the four motors could be implemented and be
more advantageous.

\section{Materials and Methods}
We have designed and implemented a system that can
provide full automatic control of a vast majority of standard
kinematic mounts available on the market. It combines open-source
resources such as 3D printing and simple electronics.

A single unit of the system, capable of controlling the tip and
tilt of a kinematic mount, is composed of two stepper motors
attached to a 3D printed plaque that aligns the knobs of the
kinematic mount with the axis of the motors. The axis are coupled
to the knobs using a 3D printed knob adapter. The motor drivers
are connected to an Arduino MEGA sensor shield (attached to and
Arduino MEGA) using Dupont wire jumper cables. Table
\ref{tab:materials} provides the list of materials required to
motorize a kinematic mount for a single channel as well as its
estimated cost.

\begin{table}[h!]
\centering
\begin{tabular}{lll}
\toprule
\textbf{Component}  & \textbf{Quantity} & \textbf{Unit Cost [ EUR ]} \\
\midrule
Parametric plaque (3D printed) & 1 & 2\\
Knob adapter (3D printed) & 2 & 2\\
M3 screw & 4 & 0.25\\
Motor 28BYJ-48 & 2 & 1.5\\
Arduino MEGA board & 1 & 10.0\\
Sensor shield for Arduino MEga & 1 & 3.0\\
Dupont cable (patch of 6 cables) & 2 & 1\\
\bottomrule
& TOTAL & 25\\
\bottomrule 
\end{tabular}
\caption{List of materials.} \label{tab:materials}
\end{table}

\subsection{Hardware}
\begin{figure}[!ht]
\centering
\includegraphics[width=14cm]{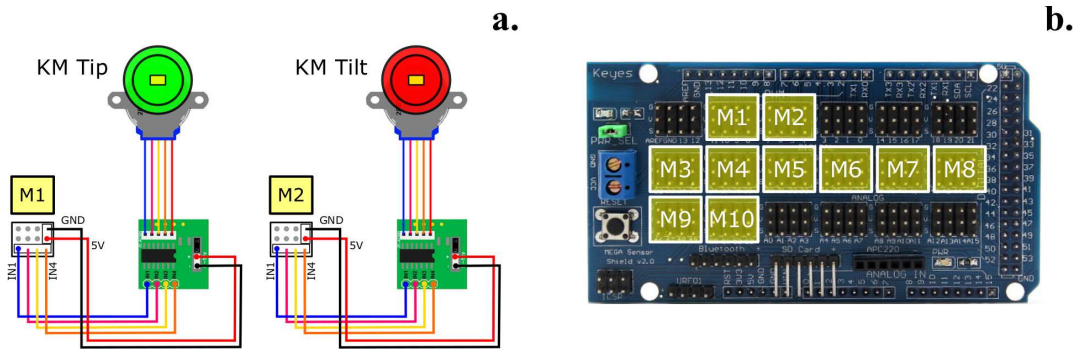}
\caption{Connection diagram for two channels.
Figure a) shows the connection diagram between
the motors and the driver electronics. Figure b)
shows the different ports available on the Arduino sensor
shield.} \label{fig:figure6}
\end{figure}

Figure \ref{fig:figure6}(a) provides a connection diagram for a
single unit that makes use of two channels (tip and tilt). Each
channel is composed of a stepper motor 28BYJ-48, driver
electronics and a Dupont cable patch with six wires. Given the
number of output pins available on the Arduino sensor shield, the
board can be used to control up to five kinematic mounts
simultaneously as shown in Figure \ref{fig:figure6}B.

\begin{figure}[!ht]
\centering
\includegraphics[width=14cm]{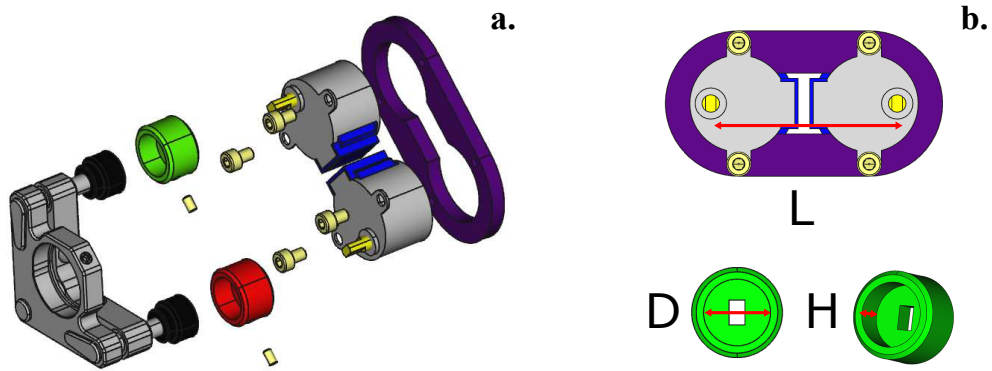}
\caption{Exploded view of the mechanical
interface. The motors are attached to the 3D printed plaque (shown
in violet) using four M4 screws. The 3D printed adapter knobs
(shown in green and red) are secured to the kinematic mount knobs
using M3 grub screws.} \label{fig:figure7}
\end{figure}

Since the 3D printed components that couple the stepper motors to
the kinematic mount were designed using FreeCAD 0.16, the distance
between motor axis on the plate, and the inner diameter and height
on the knob adapters can be easily modified using a spreadsheet.
To attach the motors to the plastic plate, four M4 screws are
required, whereas two M3 grub screws are used to secure the knobs
to the knob adapters (see figure \ref{fig:figure7}).

Table \ref{tbl:3D_parameters} provides information on the
parameters used for customizing the plate and the knob adapter to
fit into four different types of kinematic mounts (see Figure
\ref{fig:figure7}(b)). The FreeCAD, FCStd files are available on a
Thingiverse [reference].

\begin{table}[h!]
\centering
\begin{tabular}{llllll}
\toprule
\textbf{Parameter}  & \textbf{Variable} & \textbf{PLA} & \textbf{Radiant Dyes} & \textbf{Thorlabs} & \textbf{Liop-Tec}\\
\midrule
Plaque axis distance & L & 45.0 & 43.8 & 53.3 & 49.0\\
Knob diameter & D & 15.0 & 16.1 & 14.5 & 15.0\\
Knob height & H & 9.0 & 6.0 & 8.0 & 9.0\\
\bottomrule
\end{tabular}
 \caption{Parameters for 3D printed components. All
dimensions are in millimeters.} \label{tbl:3D_parameters}
\end{table}

Regarding the motors used, the 28BYJ-48 is a very cheap stepper
motor that requires five wires for its connection to the motor
driver circuit. The motor can operate in the voltage range
from 5V to 12V, and gives 4096 steps per one
turn. This number comes from the fact that the internal stepper
motor that requires 64 steps to give a complete turn is connected
to a set of gears that provide a reduction ratio of 1/64. As a
result the 28BYJ-48 requires $64\times64=4096$ steps to complete a
full turn.

\begin{figure}[!ht]
\centering
\includegraphics[width=12cm]{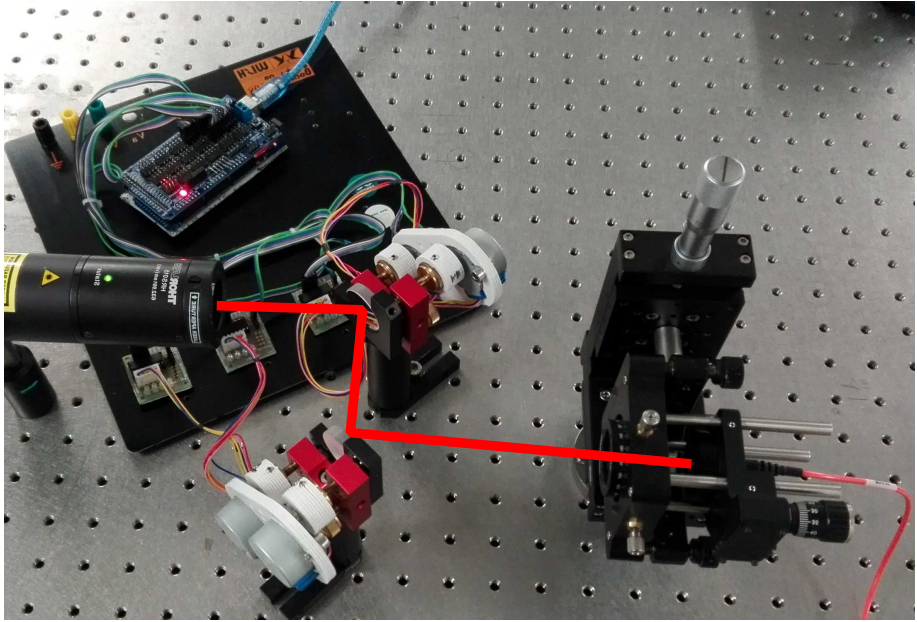}
\caption{Overall image of the system used to
improve the light coupling efficiency into a
multi-mode (MM) optical fiber.} \label{fig:figure8}
\end{figure}

For the sake of illustration, Figure \ref{fig:figure8} shows the
experimental setup corresponding to Figure \ref{fig:figure2}(b),
where two kinematic mounts from Radiant Dyes are motorized. Notice
that for each channel we have used the following components: one
3D printed supporting plate, two 3D printed knob couplers, two
stepper motors and the corresponding driving electronics. In this
particular case, the power supplied by the USB port is enough to
move the motors. However, in order to control
more than four stepper motors simultaneously, it may be necessary
to connect the sensor shield to an external power supply.

\subsection{Software}
In order to provide flexibility and simplicity from the software
perspective, each motor is controlled
independently by sending a simple instruction defined by a command
table through the serial port. A command table is a dictionary of
instructions that can be divided into two categories: commands and
queries. The first category corresponds to a direct order such as
"move stepper motor connected to port 3, half-turn". The second
type of command is used to ask information to the device about its
status; for example the state of a digital variable, the value of
a given counter or the identification string that contains
relevant information about the device and its manufacturer.

For reference, table \ref{tbl:command_table} shows the complete
list of commands used to control and retrieve information
about the status of each motor connected to the
Arduino board. Since the status of each motor is determined by a
variable that keeps track of the number of steps given by the
motor, it is assumed a one-to-one correspondence between the
number of steps defined by software and the number of steps given
by the axis of the motor. A step on the clock-wise direction adds
one unit to the global counter; a step on the counter-clock-wise
direction subtracts a unit. To keep the current status of each
motor over time, the value of each global counter is stored in the
non volatile section of the Arduino (EEPROM memory) so that all
the information is available even though the Arduino board is
powered off.

\begin{table}[h!]
\centering
\begin{tabular}{ll}
\toprule
\textbf{Command}    & \textbf{Description}\\
\midrule
STPM:N:ABS:X & move STePper Motor N to ABSolute position X (in steps)\\
STPM:N:REL:X & move STePper Motor N, RELative to current position\\
STPM:N:RST   & ReSeT STePper Motor N counter to 0\\
STPM:N:VEL:X & set STePper Motor N, VELocity to X, where X [1 fast | 10 slow]  \\
STPM:N:ST?   & retrieve motor N current STatus (position, velocity, state)     \\
             & velocity -> [1 fast | 10 slow]                                  \\
             & state -> [0 stop | 1 moving]                                    \\
STOP         & emergency stop. Press reset button to restart                   \\
\bottomrule
\end{tabular}
 \caption{List of commands and queries to control and retrieve information from the stepper motors.}
\label{tbl:command_table}
\end{table}

Since the control method is based on a command table, there are
many alternatives to implement the software interface that
controls the system. Depending on the application, the user may
require to perform simple actions driven by simple commands, or a
more elaborate sequence of tasks dependent on external signals or
defined by an algorithm with temporal dependence. The former can
be implemented by sending commands through the serial port using a
serial communication terminal compatible with the RS232 standard
such as Termite [reference] and the latter can be implemented
using MATLAB, Python or other software.

In our implementation a MATLAB routine communicates with the
Arduino board, moves the stepper motors and reads the power
detected by a Thorlabs PM100-A detector. The beam auto-alignment
procedure is carried out by scanning the X-Y plane in two steps:
firstly, the horizontal direction (X) is scanned a certain number
of steps while the optical power is measured. Afterwards, the
vertical direction (Y) is scanned.  In each step
the power and motor position are recorded and stored on a
\textit{.dat} file. When the loop finishes, the
algorithm returns the motors to the position at which maximum
power was detected. The power is first maximized
for the X direction and later one for the Y direction.

The number of steps used in each scan is selected initially by the
user. Larger ranges lead to a wider scanning of
the X-Y plane. This setting is advisable for a first approximation
while smaller rotational ranges guarantee the convergence and
optimization of the auto-alignment method implemented in MATLAB.

\section{Conclusions}
In this article we demonstrate a system that can be used to
motorize most of the kinematic mounts available
on the market. Notwithstanding being implemented using 3D printed
components, and Arduino board and simple electronics, the system
has proven to be effective, simple, flexible, and
low-cost. After performing two different
experiments where a laser beam is aligned through two reference
points, we can conclude that the system is suitable for
applications where optical realignment using flat
mirrors mounted on kinematic mounts is required on a periodic
basis. Even though the system is initially intended to motorize
kinematic mounts, its hardware and software can  be easily
extended to control other opto-mechanical components such as
translation stages.

\vspace{6pt}



\section{Author Contributions}
``L.J.S.S. and G.J. conceived and designed the experiments; L.J.S.S implemented the motorized kinematic mount system; G.J. performed the experiments; L.J.S.S., G.J. and J.P.T. analyzed the data; L.J.S.S., G.J, J.P.T. wrote the paper.

\bibliographystyle{mdpi}

\renewcommand\bibname{References}




\end{document}